\documentclass[twocolumn,prl,aps,superscriptaddress,showpacs]{revtex4}
\usepackage{graphicx}
\begin{document}

\title{Explicit illustration of non-abelian fusion rules in a small spin lattice  }

\author{Yue Yu}
\affiliation{Institute of Theoretical Physics, Chinese Academy of
Sciences, P.O. Box 2735, Beijing 100190, China}
\author{Tieyan Si}
\affiliation{Institute of Theoretical Physics, Chinese Academy of
Sciences, P.O. Box 2735, Beijing 100190, China}
\date{\today}
\begin{abstract}
We exactly solve a four-site spin model with site-dependent
Kitaev's coupling in a tetrahedron by means of an analytical
diagonalization. The non-abelian fusion rules of eigen vortex
excitations in this small lattice model are explicitly illustrated
in real space by using Pauli matrices. Comparing with solutions of
Kitaev models on large lattices, our solution gives an intuitional
picture using real space spin configurations to directly express
zero modes of Majorana fermions, non-abelian vortices and
non-abelian fusion rules. We generalize the single tetrahedron
model to a chain model of tetrahedrons on a torus and find the
non-abelian vortices become well-defined non-abelian anyons. We
believe these manifest results are very helpful to demonstrate the
nonabelian anyon in laboratory.
\end{abstract}


\maketitle

The spin lattice models of Kitaev-type have attracted many
research interests because of the abelian and nonabelian anyons in
these exactly soluble two-dimensional models
\cite{kitaev,kitaev1},  which are of the potential application in
the topological quantum memory and fault-tolerant topological
quantum computation \cite{rev}.

The abelian anyons can be explicitly shown in Kitaev toric code
model \cite{kitaev} or Levin-Wen model \cite{lw}. Fusion rules and
braiding matrix can be easily illustrated in real space. This has
simulated many attempts to design and process experiments to
demonstrate these abelian anyons in laboratory \cite{exp}.

In solving Kitaev honeycomb model \cite{kitaev} and its
ramifications \cite{yuwang,yao1,wuc}, a key technique is the usage
of the Majorana fermion representation of the spin-1/2 operators.
The systems then are transferred into bilinear fermion systems and
the ground state sector can be diagonalized in the momentum space.
The shortcoming to use the momentum space is that the ground state
and the elementary excitations are hard to be expressed by the
original spin operators, i.e., Pauli matrices. On the other hand,
the sectors with vortex excitations can only be treated
numerically. Then the nonabelian fusion rules and statistics may
not be directly shown in Pauli matrices' language. Lahtinen et al
have derived the nonabelian fusions through the spectrum analysis
\cite{la}. However, they are still not directly related to Pauli
matrices. A real space form of the ground state of the Kitaev
honeycomb model in the insulating phase with abelian anyon was
studied \cite{cn}. We tried to present the non-abelian fusion
rules for high energy excitations in this Kitaev model
\cite{yusi}. There was an attempt to use toric code abelian anyons
to superpose the Ising non-abelian anyons without involving a
Hamiltonian\cite{wang}.

\begin{figure}[htb]
\begin{center}
\includegraphics[width=5cm]{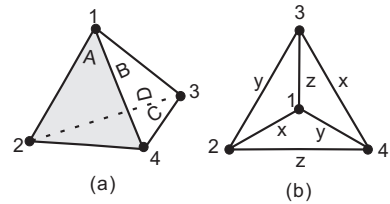}
\end{center}
\caption{\label{fig:Fig.1} (a) A tetrahedron in whose points 1,2,3
and 4, the spins live. The four surfaces are labelled by $A, B, C$
and $ D$ or 124, 134, 234 and 123. (b) The top view of the
tetrahedron. $x,y,z$ are the links with different Kitaev
couplings.}
\end{figure}

In this paper, we solve a spin model with Kitaev's coupling in a
small lattice, i.e., a tetrahedron (Fig.\ref{fig:Fig.1}). We
perform Majorana fermions and their zero modes, non-abelian
vortices and their fusion rules in real space by means of Pauli
matrices. Because the system is finite and everything can be
deduced in an elemental way, it will be very helpful to
intuitionally understand these concepts which were used to be
expressed in those deep mathematical language. Since the spin
configurations of these excitations and the fusion rules are
explicitly shown, the experimental techniques with cold atoms,
finite photon graph states and nuclear magnetic resonance systems
\cite{exp} are possible to be applied to demonstrate these quantum
states and then to design quantum bits and gates for a topological
quantum computer.

The single tetrahedron model is too small for non-abelian anyons
to be well-defined. We generalize it to be a chain model of
tetrahedrons on a torus. This chain model is also exactly solvable
and the non-abelian anyons can be well-defined. They are still of
a simple form like the non-abelian vortex in the single
tetrahedron model.

\vspace{2mm}

\noindent{\it The model and symmetries } Kitaev model in a
tetrahedron  is given by $ H_K=\sum_{x}J_{x,ij}\sigma_i^x
\sigma_j^x+\sum_{y}J_{y,ij}\sigma_i^y
\sigma_j^y+\sum_{z}J_{z,ij}\sigma_i^z \sigma_j^z$  where $x,y,z$
are the links shown in Fig. \ref{fig:Fig.1}(b) and $J_{a,ij}$ are
link- and site-dependent coupling constants. $\sigma^a_j$ are
spin-1/2 operators obeying Pauli matrix algebra, e.g.,
$\sigma^x_i\sigma^y_i=i\sigma_i^z$ and
$(\sigma^x_i)^2=(\sigma^y_i)^2=(\sigma^z_i)^2=1$. The spin
operators on the different sites commute, i.e.,
$[\sigma^a_i,\sigma^b_j]=0$ for $i\ne j$.  An intuitive
imagination is the model may be easily solved if $J_{a,ij}$ are
not site-dependent as that in Kitaev model in an infinite lattice
\cite{kitaev1}. However, a direct check finds that, unlike Kitaev
model in an infinite lattice, this model can not be reduced to a
bilinear fermion theory in such a coupling constant choice. This
is because the tetrahedron is topologically equivalent to a
sphere, Kitaev model defined on this compact space is very
different from the model on infinite or periodic lattices. In this
paper, we consider these coupling constants are site-dependent. We
also include some three- and four-spin coupling terms as those in
a generalized Kitaev model \cite{yuwang}. The model Hamiltonian we
will study is given by
\begin{eqnarray}
&&H=J_x\sigma_1^x\sigma^x_2+J_x\sigma_3^x\sigma^x_4
+J_y\sigma^y_2\sigma^y_3 \nonumber\\&&+\kappa
\sigma_1^x\sigma_2^z\sigma_3^y+\kappa\sigma_2^y\sigma_3^z\sigma_4^x
+\lambda\sigma_1^x\sigma_2^z\sigma_3^z\sigma_4^x.\label{1}
\end{eqnarray}
There are four conserved operators which live in triangular
plaquettes:
\begin{eqnarray}
&&P=\{P_A=P_{124}=\sigma^z_1\sigma_2^y\sigma^x_4,~
P_B=P_{134}=\sigma^x_1\sigma_3^y\sigma^z_4,\nonumber\\&&P_C=P_{234}=\sigma^x_2\sigma_3^z\sigma^y_4,~P_D=P_{123}
=\sigma^y_1\sigma_2^z\sigma^x_3 \},
\end{eqnarray}
which are mutual commutative. They obey $[P_{ijk},H]=0$ and
$P_AP_BP_CP_D=1$. Three-spin couplings break time reversal
symmetry. The Hamiltonian is time-reversal invariant if
$\kappa=0$, .

\vspace{2mm}

\noindent{\it Bilinearization, diagonalization and states } This
Hamiltonian can be transferred into a bilinear fermion
Hamiltonian. In the deduction process, we can illustrate some
abstract concepts in an elementary way. For example, we can define
eight Majorana fermions corresponding to four sites:
\begin{eqnarray}
&&\psi_1=\sigma_1^y,\psi_2=\sigma^x_2\sigma^z_1,
\psi_3=\sigma^y_3\sigma_2^z\sigma^z_1,
\psi_4=\sigma^x_4\sigma^z_3\sigma_2^z\sigma^z_1 \label{3}\\
&&b_1=-\sigma_1^x,b_2=-\sigma^y_2\sigma^z_1,
b_3=-\sigma^x_3\sigma_2^z\sigma^z_1,
b_4=-\sigma^y_4\sigma^z_3\sigma_2^z\sigma^z_1\nonumber
\end{eqnarray}
They obey $\{\psi_i,\psi_j\}=\{b_i,b_j\}=2\delta_{ij}$ and
$\{\psi_i,b_j\}=0$. Remarkably, $[H,b_i]=0$. This can be directly
checked or be seen by writing the Hamiltonian in terms of Majorana
fermions,
\begin{eqnarray}
H=-t(d^\dag_a d_b+d^\dag_b d_a)-\mu(d^\dag_ad_a+ d^\dag_bd_b)
+\Delta^* d_ad_b+\Delta d_b^\dag d^\dag_a,\label{2}
\end{eqnarray}
where $ d_a=\frac{1}2(\psi_1+i\psi_2)$ and
$d_b=\frac{1}2(\psi_3+i\psi_4)$. The parameter relations are given
by
 $\mu=2J_x,~t=J_y+\lambda,~\Delta=\Delta_1+i\Delta_2
 =J_y-\lambda+i\kappa.$ An easy way to identify (\ref{1}) and
 (\ref{2}) is substituting (\ref{3}) into (\ref{2}). One can check
 $d_a$ and $d_b$ are conventional fermions, i.e.,
 $\{d_a,d^\dag_a\}=\{d_b,d^\dag_b\}=1$ and
 $\{d_a,d_a\}=\{d_b,d_b\}=\{d^\dag_a,d_b\}=\{d_a,d^\dag_b\}=0$ \cite{cn,yuwang}. This is
 a BCS $p$-wave pairing Hamiltonian in a finite system and can be
 diagonalized by rewriting (\ref{2}) as
\begin{equation}
H=\frac{1}2(d_a^\dag,d_b^\dag,d_a,d_b)\left(\begin{array}{cccc}
    -\mu&-t&0&-\Delta\\
    -t&-\mu&\Delta&0\\
     0&\Delta^*&\mu&t\\
    -\Delta^*&0&t&\mu\\\end{array}\right)\left(\begin{array}{c}
    d_a\\
    d_b\\
    d^\dag_a\\
    d^\dag_b\\\end{array}\right)
\end{equation}
The eigen vaules of the Hamiltonian matrix are
\begin{eqnarray}
&&E_0=-\frac{1}2(\sqrt{|\Delta|^2+\mu^2}+t),
E_t=-\frac{1}2(\sqrt{|\Delta|^2+\mu^2}-t),\nonumber\\
&&E_\mu=\frac{1}2(\sqrt{|\Delta|^2+\mu^2}-t),
E_2=\frac{1}2(\sqrt{|\Delta|^2+\mu^2}+t),
\end{eqnarray}
Diagonalizing the Hamiltonian, one has
\begin{eqnarray}
H&=&E_2\tilde d^\dag_b\tilde d_b+E_\mu\tilde d^\dag_a\tilde
d_a+E_0\tilde d_b\tilde d^\dag_b+E_t\tilde d_a\tilde d^\dag_a.
\end{eqnarray}
The generalized Bogoliubov fermions $\tilde d_{a,b}$ can be
obtained in a standard way with $\tilde
d_{a,b}=x^{(a,b)}_1d_{a,b}+x^{(a,b)}_2d_{b,a}+x^{(a,b)}_3d^\dag_{a,b}+x^{(a,b)}_4d^\dag_{b,a}
$ and the coefficients $x^{a,b}_i$ are normalized eigen vector of
the Hamiltonian matrix. The Bogoliubov fermion operators obey the
standard fermion commutation relations. A subspace of quantum
states are $\{|G\rangle,\tilde d_a|G\rangle,\tilde
d^\dag_a|G\rangle, \tilde d^\dag_b\tilde d^\dag_a|G\rangle\}$
where  $|G\rangle=\tilde d_a\tilde d_b|0\rangle$ with $
d_a|0\rangle= d_b|0\rangle=0$. The vacuum
$|0\rangle=d_ad_b|\phi\rangle$ for  a reference state
$|\phi\rangle$, e.g., $|$$\uparrow\uparrow\uparrow\uparrow
\rangle$. The eigen energies of this set of quantum states are
$\{-\sqrt{|\Delta|^2+\mu^2},-t,t,\sqrt{|\Delta|^2+\mu^2}\}$. When
$\sqrt{|\Delta|^2+\mu^2}>t$, i.e.,
$4J_x^2+\kappa^2-4J_y\lambda>0,$ $|G\rangle$ is the ground state.
Because $[H,b_i]=0$, each energy level is formally sixteen-fold
degenerate, e.g., the ground states are $ \{|G\rangle,
b_i|G\rangle,b_ib_j|G\rangle,b_ib_jb_k|G\rangle\}. $ That is ,
$b_i$ play a role of zero modes of Majorana fermions. However,
there are only four independent, which, e.g., are
\begin{eqnarray}
\{|G\rangle,c_1^\dag|G\rangle,c_2^\dag|G\rangle, c_1^\dag
c_2^\dag|G\rangle\}\label{ground}
\end{eqnarray}
where $c_1^\dag=\frac{1}2(b_1-ib_3)$ and
$c^\dag_2=\frac{1}2(b_2-ib_4)$ with  $
c_1|G\rangle=c_2|G\rangle=0$.
 The total Hilbert space is
sixteen-dimensional as expected. Four degenerate states in a given
energy level are distinguished by quantum number
$P=(P_A,P_B,P_C,P_D)$, which are shown in Tab. 1.

\vspace{3mm}

\def\arraystretch{0}\begin{tabular}{|c|c|c|c|c|}
\hline\small
 &$1$&$c^\dag_1$&$c^\dag_2$&$c^\dag_1c^\dag_2$\\\hline
$|G\rangle$&(1,1,1,1)&(-1,1,1-1)&(1,-1,-1,1)&(-1,-1,-1,-1)\\\hline
$\tilde
d^\dag_a|G\rangle$&(-1,-1,1,1)&(1,-1,1-1)&(-1,1-1,1)&(1,1,-1,-1)\\\hline
$\tilde
d^\dag_b|G\rangle$&(-1,-1,1,1)&(1,-1,1-1)&(-1,1-1,1)&(1,1,-1,-1)\\\hline
$\tilde d^\dag_a\tilde
d^\dag_b||G\rangle$&(1,1,1,1)&(-1,1,1-1)&(1,-1,-1,1)&(-1,-1,-1,-1)\\\hline
 \end{tabular}

\vspace{3mm}

{\small Table 1: The eigen values of $P$ of the quantum states.}

\vspace{3mm}

Due to the constraint $P_AP_BP_CP_D=1$, there are eight different
$P$ which are carried by the states in the first two levels or in
the last two levels as shwon in Tab. 1.

\vspace{2mm}

\noindent{\it Fusion rules: abelian and non-abelian } We now go to
illustrate the fusion rules of these eigen excitations. First, we
define Majorana fermions corresponding to $\tilde d_{a,b}$:
\begin{eqnarray}
&&\tilde \psi_1=\tilde d_a+\tilde d^\dag_a,~\tilde
\psi_2=-i(\tilde d_a-\tilde d^\dag_a),\nonumber\\&&
\tilde
\psi_3=\tilde d_b+\tilde d^\dag_b,~\tilde \psi_4=-i(\tilde
d_b-\tilde d^\dag_b).
\end{eqnarray}
They obey $\{\tilde\psi_i,\tilde\psi_j\}=2\delta_{ij}$ and
$\{\tilde\psi_i,b_j\}=0$. There are four sets of states which obey
abelian fusion rules, as those in Kitaev toric code model,
\begin{eqnarray}
&&\sigma^{(1)}_i\sigma^{(2)}_i\sim \tilde \psi_i,~
\sigma^{(1)}_i\tilde \psi_i\sim \sigma^{(2)}_i,~
\sigma^{(2)}_i\tilde \psi_i\sim \sigma^{(1)}_i,\nonumber
\\&&
\sigma^{(1)}_i\sigma^{(3)}_i\sim b_i,~~\sigma^{(1)}_ib_i\sim
\sigma^{(3)}_i,~~ \sigma^{(3)}_ib_i\sim \sigma^{(1)}_i,
\end{eqnarray}
for $i=1,2,3,4$. These operators are
\begin{eqnarray}
&& \sigma^{(1)}_1=ib_1\tilde \psi_1,~\sigma^{(2)}_1=ib_1,
~ \sigma^{(3)}_1=-i\tilde \psi_1;\nonumber\\
&&\sigma^{(1)}_2=-ib_2\tilde \psi_2,~\sigma^{(2)}_2=-ib_2,~
\sigma^{(3)}_2=i\tilde \psi_2;
\\
&& \sigma^{(1)}_3=-b_1\tilde \psi_1b_3\tilde \psi_3,~
\sigma^{(2)}_3=-b_3b_1\tilde \psi_1,~\sigma^{(3)}_3
=\tilde \psi_3b_1\tilde \psi_1;\nonumber\\
&&\sigma^{(1)}_4=-b_2\tilde \psi_2b_4\tilde \psi_4,~
\sigma^{(2)}_4=-b_4b_2\tilde \psi_2, ~\sigma^{(3)}_4 =-\tilde
\psi_4b_2\tilde \psi_2.\nonumber
\end{eqnarray}
 Acting on the ground state
$|G\rangle$, they are eigen states of $P$ and their eigen values
can be read out from Tab. 1. They are also eigen states of the
Hamiltonian and their eigen energies can be read out from the
number of $\tilde\psi_i$ in a given operator. Since each energy
level is four-fold degenerate, we find that the linear combination
of these degenerate states may obey non-abelian fusion rules. They
can be thought as non-abelian vortices. For example,
\begin{eqnarray}
&&\eta_A=\frac{i}{\sqrt 2} (b_2\tilde \psi_2-b_4b_2\tilde
\psi_2),~\eta_B=\frac{i}{\sqrt
2}(b_1\tilde\psi_1-b_3b_1\tilde\psi_1),\nonumber\\
&&\eta_C=\frac{i}{\sqrt
2}(b_4\tilde\psi_4-b_1b_4\tilde\psi_4),~\eta_D=\frac{i}{\sqrt
2}(b_3\tilde\psi_3-b_2b_3\tilde\psi_3).
\end{eqnarray}
They are in fact the superposition of those toric code abelian
vortcies. (We will be back to this point when we study braiding of
anyons.) Acting these operators on $|G\rangle$, they are also
eigen states of $H$. The details of $P$ and $H$'s eigen values
list on Tab. 2. The subscript indices are used because, e.g.,
$\eta_A|G\rangle$ has an eigen value $P_A=-1$ and other three
either are 1 or do not have a definite eigen value. Therefore,
$\eta_A$ can be thought as a vortex excitation on the surface $A$,
and so on.

 \vspace{3mm}

\hspace{0.7cm}\def\arraystretch{0}\begin{tabular}{|c|c|c|c|c|c|}
\hline
 &$~~P_A~~$&$~~P_B~~$&$~~P_C~~$&$~~P_D~~$&$~~H~~$\\\hline
$~~\eta_A~~$&-1&*&*&1&$-t$\\\hline $\eta_B$&*&-1&1&*&$-t$\\\hline
$\eta_C$&*&1&-1&*&$t$\\\hline $\eta_D$&1&*&*&-1&$t$\\\hline
 \end{tabular}

\vspace{2mm}

\noindent{\small Table 2: The eigen values of $P$ and $H$ of
$\eta$. `$*$' means the vortex does not have a definite eigen
value. }

\vspace{2mm}

Using the algebra of Pauli matrices or equivalently, the
anti-commutation relations between the Majorana fermions, one may
directly prove that these operators obey the following non-abelian
fusion rules
\begin{eqnarray}
&&\eta_A\eta_A\sim 1+b_4, ~~\eta_A b_4\sim \eta_A,~~
b_4b_4=1,\nonumber\\&&\eta_B\eta_B\sim 1+b_3, ~~\eta_B b_3\sim
\eta_B,~~
b_3b_3=1,\label{naf}\\
&&\eta_C\eta_C\sim 1+b_1, ~~\eta_C b_1\sim \eta_C,~~
b_1b_1=1,\nonumber\\ &&\eta_D\eta_D\sim 1+b_2, ~~\eta_D b_2\sim
\eta_D,~~ b_2b_2=1,\nonumber
\end{eqnarray}
which are standard non-abelian Ising fusion rules. Equations
(\ref{naf}) are one of central results in this paper. Only when
the zero modes of Majorana fermions exist, the non-abelian
vortices are eigen excitations \cite{iv}. We hope this
illustration can help condensed matter physicists have a direct
impression to these elusive mathematical relations and understand
them in an elementary way.

\vspace{2mm}

\noindent{\it Breaking of time reversal symmetry } The Ising
non-abelian fusion rules and time reversal symmetry are
concomitant. The three-spin coupling terms in Hamiltonian
(\ref{1}) explicitly break the time reversal symmetry. However,
our deduction of the non-abelian fusion rules does not rely on a
non-zero $\kappa$. They hold even $\kappa=0$. The only change is
$\Delta_2=\kappa=0$ and the gap $\Delta=\Delta_1$. Actually, when
$\kappa=0$, the time reversal symmetry is spontaneously broken.
The ground states are four-fold degenerate, which are given by
(\ref{ground}). Since under the time reversion $T$, $T\sigma^a
T^{-1}=-\sigma^a$, one has $T|G\rangle=|G\rangle,~
Tc^\dag_2|G\rangle=c^\dag_2|G\rangle,
~Tc_1^\dag|G\rangle=-c_1^\dag|G\rangle,~Tc_1^\dag
c^\dag_2|G\rangle=-c_1^\dag c^\dag_2|G\rangle.$ We see two sectors
which have different eigen values of  $T$, i.e., the time reversal
symmetry is spontaneously broken. This is because of the geometric
frustration of the tetrahedron. This spontaneous breaking of the
time reversal symmetry first observed in Kitaev model on a
triangle-honeycomb lattice and leads to a chiral spin liquid
\cite{yao}.

\vspace{2mm}

\noindent{\it Non-abelian anyons } A frequently quoted result is
that the excitations obey non-abelian fusion rules like
(\ref{naf}) is equivalent to the braidings of these vortices
$\eta$ are non-abelian  and then these vortices are called anyons
with non-abelian statistics or non-abelian anyons \cite{rev}.
However, it is based on these anyons are well-defined and they are
energetically degenerate. In this small system, to identify an
$\eta$ vortex as an anyon is reluctant because we see that, e.g.,
$\eta_A$ is not an eigen state of $P_B$ and $P_C$, which means
this is not a particle-like isolated excitation. On the other
hand, the vortices $\eta_{A,B}$ and $\eta_{C,D}$ are not in the
same energy level if $t\ne 0$ and braiding two vortices with
different energy do not make sense in statistics. Therefore, for
this small system, we only emphasize the non-abelian fusion rules
of these vortex excitations but not call them non-abelian anyons.

\begin{figure}[htb]
\begin{center}
\includegraphics[width=2.3cm]{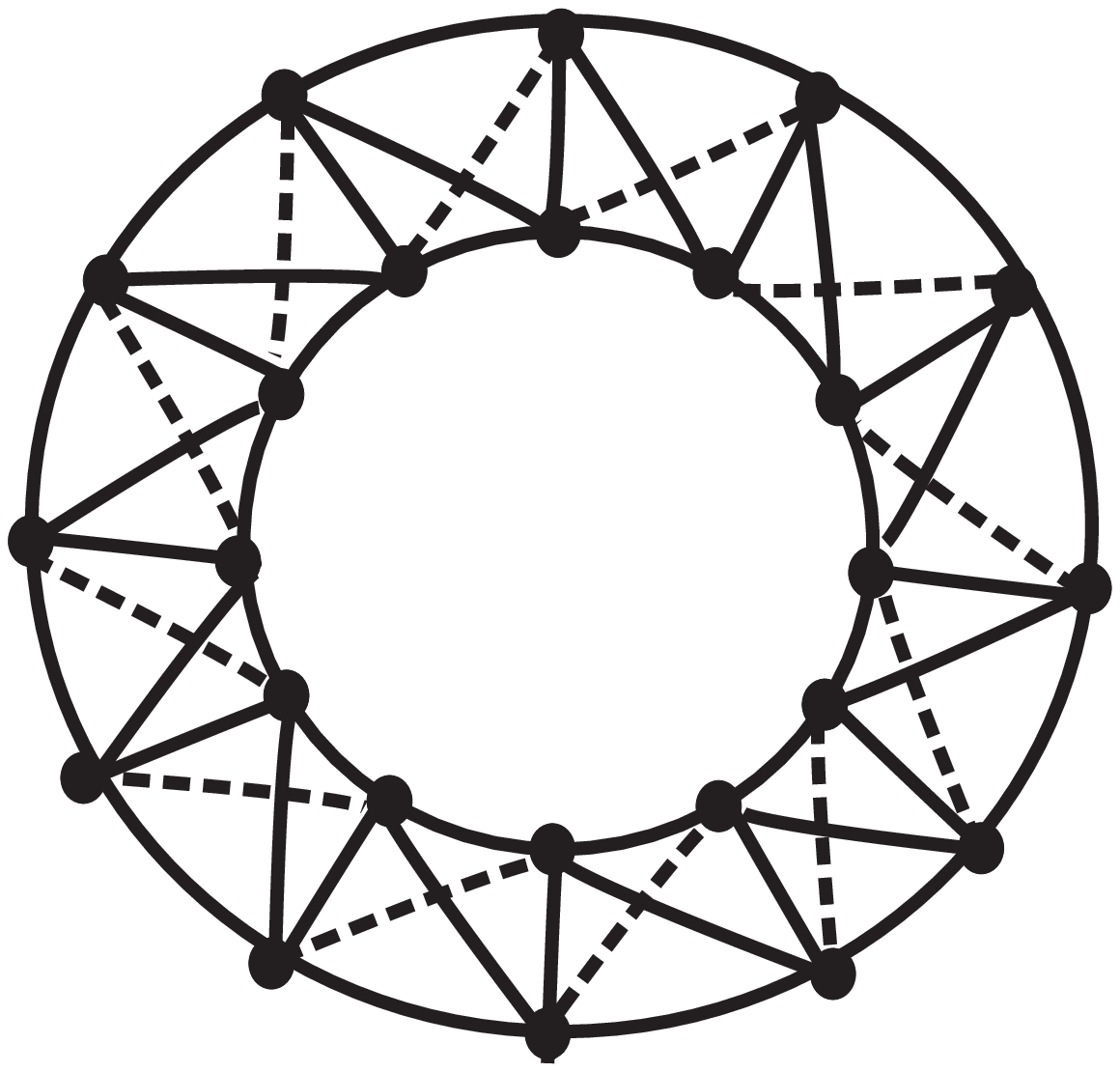}
\includegraphics[width=6.5cm]{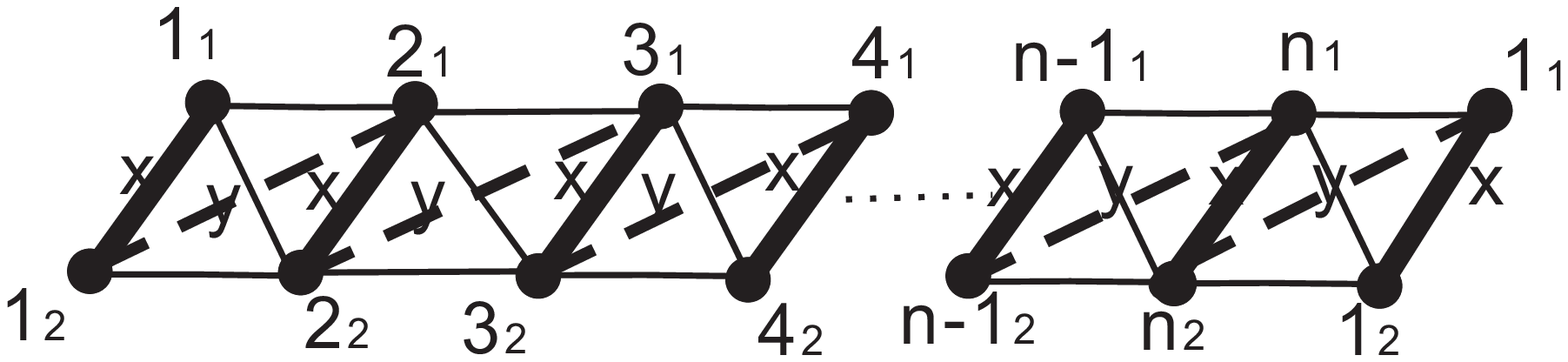}
\end{center}
\caption{\label{fig:Fig.2} left panel: A chain of tetrahedrons on
a torus. Right panel: unwind the torus to a periodic lattice. The
thick lines carry $J_x$ while the dash lines carry $J_y$.}
\end{figure}

To well define an anyon, we generalize the  single tetrahedron
model to a chain model of tetrahedrons on a torus (Fig.
\ref{fig:Fig.2}). The model Hamiltonian is given by
\begin{eqnarray}
&&H_{\rm chain}=
J_x\sum_{i=1}^{n}\sigma^x_{i_1}\sigma^x_{i_2}+J_y\sum_{i=1}^{n}
\sigma^y_{i_2}\sigma^y_{i+1_1}\nonumber\\
&&\label{gh}
+\lambda\sum_{i=1}^{n}\sigma^x_{i_1}\sigma^z_{i_2}\sigma^z_{i+1_1}
\sigma^x_{i+1_2}\\
&&
+\kappa\sum_{i=1}^{n}(\sigma^x_{i_1}\sigma^z_{i_2}\sigma^y_{i+1_1}
+\sigma^y_{i_2}\sigma^z_{i+1_1}\sigma^x_{i+1_2})\nonumber
\end{eqnarray}
with $n+1\equiv 1$. Only a half of triangular plaquette operators
are conserved, which are $
P=\{P_{2a-1}=P_{2a-1_12a-1_22a_1}=\sigma^y_{2a-1_1}\sigma^z_{2a-1_2}\sigma^x_{2a_1},
P_{2a}=P_{2a-1_12a_12a_2}=\sigma^y_{2a-1_2}\sigma^z_{2a_1}\sigma^x_{2a_2}\}.$
The Hamiltonian can be diagonalized as
\begin{eqnarray}
&&H_{\rm chain}=-t\sum_{i=1}^{n}(d^\dag_i d_{i+1}+d^\dag_{i+1}
d_i)-\mu\sum_{i=1}^{n}d^\dag_id_i\nonumber\\&&+\sum_{i=1}^{n}(\Delta^*
d_id_{i+1}+\Delta d_{i+1}^\dag d^\dag_i),
\end{eqnarray}
where  $ d_i=\frac{1}2(\psi_{i_1}-i\psi_{i_2}) $ and the Majorna
fermions are given by \cite{jw}\cite{cn}
$\psi_{i_1}=\sigma^x_{i_1}\prod_{j<i}\sigma^z_{j_1}\sigma^z_{j_1},
\psi_{i_2}=\sigma^y_{i_2}\sigma^z_{i_1}\prod_{j<i}\sigma^z_{j_1}\sigma^z_{j_1},~b_{i_1}=-\sigma^y_{i_1}\prod_{j<i}\sigma^z_{j_1}\sigma^z_{j_1},
b_{i_2}=-\sigma^x_{i_2}\sigma^z_{i_1}\prod_{j<i}\sigma^z_{j_1}\sigma^z_{j_1},
$
where Majorana fermions $b_i$ commute with $H_{\rm chain}$.
Therefore, $b_ib_j\cdots|\rangle$ are degenerate states if
$|\rangle$ is an eigen state. Since $[P,\psi_i]=0$ for all $i$ but
not all $b_i$, the eigen values of $P$ for the vortex states
$\eta_i=\frac{i}{\sqrt 2}(b_i\tilde\psi_p-b_jb_i\tilde\psi_p)$,
$j\ne i$ are determined by $b_{i,j}$. Here $\tilde\psi_p$ are the
linear combination of $\psi_l$ after diagonalizing $H_{\rm chain}$
as those in the single tetrahedron model. If $j$ is far from $i$,
$\eta_i$ has only one minus $P$ near the $i$-th site. Then,
$\eta_i$ is a well-defined single vortex operator and can be
thought as a non-abelian anyon due to  $\eta_i\eta_i=1+b_j$. There
are $n$ such anyons which are degenerate. Each anyon brings a zero
mode $b_i$ of Majorana fermion in its center
\cite{iv}\cite{kitaev}.

\vspace{2mm}

\noindent{\it Non-abelian braiding matrices} Rewriting
$\eta_i=\frac{1}{\sqrt{2}}(e_i-m_i)$ with $e_i=ib_i\tilde\psi_p$
and $m_i=ib_jb_i\tilde\psi_p$, one has abelian fusuion rules
$em\sim b$, $mb\sim e$, $eb\sim m$ and $e^2=m^2=b^2=1$. $e$ and
$m$ are toric code mutual abelian anyons with degenerate energy.
Thus, the Ising anyon is in fact the superposition of the toric
code abelian anyons. This result has been recognized in refs.
\cite{yusi,wang} but without involving the Hamiltonian. Therefore,
the toric code braiding matrices determine the non-abelian
braiding matrices of $\eta_i$ \cite{wang}, which are Ising-like
braiding matrices
$$
R^{bb}_1=-1,~(R^{\eta\eta}_1)^2=1,~ (R^{b\eta}_\eta)^2=-1,
~(R^{\eta\eta}_b)^2=-1. $$Missing of the complex phases
$e^{-i\pi/8}$ in $R^{\eta \eta}_1$ and $e^{i\pi/8}$ in
$R^{\eta\eta}_b$ is because $R=R^\dag$ for the toric code anyons
\cite{wang}. A framing therefore is  needed \cite{lw}. We do not
intend to propose a framing and ancillary qubits to implement the
non-trivial chirality but refer to Wootton et al \cite{wang}.

In conclusions, we presented a simple model in which there are a
set of vortices obeying non-abelian fusion rules which were
explicitly illustrated in an elementary way without using deep
mathematical tools. Finally, we generalized the single tetrahedron
model to a chain model of tetrahedrons on a torus and showed that
the non-abelian vortices defined in the single tetrahedron model
become well-defined non-abelian anyons.

This work was supported in part by the national natural science
foundation of China, the national program for basic research of
MOST of China and a fund from CAS.

\end{document}